\documentclass{nature}
\usepackage{graphicx}
\usepackage{float}
\usepackage{placeins}
\usepackage{gensymb}
\usepackage{bm}
\usepackage{amssymb}
\usepackage{amsmath}
\usepackage{amsmath}
\usepackage{bm}
\usepackage{lineno}
\usepackage{xcolor}
\usepackage{soul}

\usepackage{times}
\usepackage{color}
\usepackage{float}

\usepackage{braket}
\usepackage{epstopdf}
\usepackage{mathtools}
\usepackage{mathrsfs}

\usepackage{lipsum}
\usepackage{ccaption}

\usepackage{caption}

\usepackage{subfig}

\allowdisplaybreaks

\title{Dynamics of interacting fermions under spin-orbit coupling in an optical lattice clock}

\author{S. L. Bromley$^{1,*}$,
S. Kolkowitz$^{1,*}$,
T. Bothwell$^{1}$,
D. Kedar$^{1}$,
A. Safavi-Naini$^{1}$,
M. L. Wall$^{1, \dagger}$,
C. Salomon$^{2}$,
A. M. Rey$^{1}$,
J. Ye$^{1}$\\
$^*$ These authors contributed equally.}

\begin{document}

\maketitle

\begin{affiliations}
 \item JILA, NIST and Department of Physics, University of Colorado, 440 UCB, Boulder, Colorado 80309, USA
 \item Laboratoire Kastler Brossel, CNRS, UPMC, \' Ecole Normale Sup\' erieure, 24 rue Lhomond, 75231 Paris, France. \\
$^{\dagger}$Present address: The Johns Hopkins University Applied Physics Laboratory, Laurel, MD 20723, USA\\
Correspondence and requests for materials should be addressed to Sarah.L.Bromley@colorado.edu.
\end{affiliations}

\begin{abstract}

Quantum statistics and symmetrization dictate that identical fermions do not interact via $s$-wave collisions. However, in the presence of spin-orbit coupling (SOC), fermions prepared in identical internal states with distinct momenta become distinguishable.  The resulting strongly interacting system can exhibit exotic topological and pairing behaviors\cite{Qi2011,Barbarino2016,Strinati2016,Zeng2015}, many of which are yet to be observed in condensed matter systems. Ultracold atomic gases offer a promising pathway for simulating these rich phenomena\cite{ZhaiSOCReview,SpielmanReview, LewensteinReview,galitskiSOCReview}. Two recent experiments reported the observation of single atom SOC in optical lattice clocks (OLCs) based on alkaline-earth atoms\cite{Kolkowitz2017, FallaniNew}.  In these works encoding the effective spin degree of freedom in the long-lived electronic clock states significantly reduced the detrimental effects of spontaneous emission and heating that have thus far hindered the study of interacting SOC with alkali atoms\cite{DalibardReview,WallSOC}. 
Beyond first studies of interacting SOC with alkali atoms in a bulk gas\cite{ZhangFuPRA2013}, in a lattice shaken BEC\cite{ChinSOC}, and with two particles in a lattice\cite{GreinerSOC2016}, here we enter a new regime of many-body interacting SOC in a fermionic OLC.  Using clock spectroscopy, we observe the precession of the collective magnetization and the emergence of spin locking effects arising from an interplay between $p$-wave and SOC-induced exchange interactions. The many-body dynamics are well captured by a collective XXZ spin model, which describes a broad class of condensed matter systems ranging from superconductors to quantum magnets.  Furthermore, our work will aid in the design of next-generation OLCs by offering a route for avoiding the observed large density shifts caused by SOC-induced exchange interactions.

\end{abstract}

In our one-dimensional optical lattice clock (OLC) many-body effects arise from the cooperation and competition between $p$-wave and $s$-wave interactions, along with single-particle SOC dynamics. The spin-motion coupling we engineer in the OLC primarily affects how spins interact with each other, without any thermalization effects in the lattice. This unique condition sets up an effective spin system that provides a simpler view of the complex interplay between SOC and many-body interactions. Meanwhile, it grants us immediate access to quantum magnetism at $\mu$K motional temperatures.

The many-body dynamics are described by a collective XXZ spin model\cite{MJMManyBodySr,XiboSUN}, which contains both exchange ($s$- and $p$-wave)  and Ising ($p$-wave) terms. The dynamics of collective XXZ models have largely been studied theoretically in condensed matter physics, for example in the context of superconductivity through the Anderson pseudospin mapping\cite{Anderson1958}, which identifies Cooper pairs and holes as the two components of an effective pseudospin. Only limited experimental studies have been conducted so far, and they have been restricted mainly to weak quenches\cite{QuenchScience}. The ultra-narrow clock transition in our OLC enables the preparation, control, and spectroscopic resolution of the dynamics in a broad parameter space, including quenches over a large dynamic range.

SOC with strong interactions between a pair of atoms has been realised  in a lattice\cite{GreinerSOC2016}.  Here, we instead use a large atom number, $N$, to tune the strength of the interactions to enter a strong, collective interacting regime well beyond single-particle SOC dynamics.
We observe that both $s$-wave and $p$-wave interactions induce precession of the collective magnetization. Furthermore, the exchange interactions compete with the SOC-induced dephasing and promote spin alignment and locking. Similar interaction-induced spin locking effects have been observed in other trapped gas experiments\cite{LongClockCoherence,Du2008}, and were recently shown to play a crucial role in the stabilization of time crystal phases in trapped ions\cite{Zhang2017} and impurity centers in diamond\cite{Choi2017}. In those cases, however, dephasing arose from spatial inhomogeneities, in contrast to our system where dephasing is a direct consequence of an intrinsic modification of the band structure by SOC.

In our experiment, up to $\sim1.5\times10^{4}$ atoms are laser cooled into a horizontal one-dimensional (1D) lattice  operated at a ``magic'' wavelength, $\lambda_{L}=813$ nm, where the band structures of the two clock states are identical.  By changing only the retro-reflected power while fixing the incident power we keep the radial confinement approximately constant ($\nu_{r}\approx500$ Hz) while significantly modifying the tunneling rate\cite{Kolkowitz2017}. We create an array of $\sim$10$^{3}$ pancake-shaped lattice sites, each with 1 - 20 atoms when we vary $N$.

Axially the atoms occupy the ground band of the lattice. Radially the atoms are only weakly confined with a thermal distribution among the radial modes, $\mathbf{n_{r}}$.  In the tight binding limit we can write the energies of the ground bands as $E\left (  q,{\bf n}_{{\bf r}}\right )=-2\hbar J_{\mathbf{n_{r}}}\cos(q)$, where $\hbar$ is the reduced Planck constant, $J_{\mathbf{n_{r}}}$ is the tunneling rate between nearest-neighbor lattice sites that has some  dependence on the radial mode  ${\mathbf{n_{r}}}$ (see Methods), and $q$ is the quasimomentum in units of $\hbar/a$, where $a=\lambda_{L}/2$ is the lattice spacing. Due to the temperature of the atoms, the band is thermally filled and all $q$ are initially occupied.

The clock laser ($\lambda_{c}=698$ nm), aligned axially along the 1D lattice, drives transitions between the ground $^{1}$S$_{0}$ $(\ket{g, q}_{{\bf n}_{{\bf r}}})$ and the excited $^{3}$P$_{0}$ $(\ket{e, q+\phi}_{{ \bf {n}}_{{\bf r}}})$ clock states. The quasimomentum shift of the excited state, $\phi=\pi\lambda_{L}/\lambda_{c}$,  which is needed for conservation of momentum, generates the SOC. The consequence of this shift becomes important only when atoms are allowed to tunnel.  For N atoms evolving independently under SOC, the Hamiltonian can be expressed in terms of a synthetic magnetic field\cite{WallSOC,Kolkowitz2017},
\begin{equation}
\label{eq:SOCHamiltonian}
	\hat{H}_{SOC}/\hbar=-\sum_{i=1}^N \vec{B}_{SOC}(q_{i}, {\bf n}_{{\bf r}_i},\Omega,\delta) \cdot {\bf \hat S}_{i},
\end{equation}
\noindent where $\delta$ and $\Omega$ are the clock laser detuning from the bare atomic transition and  Rabi frequency, respectively, and
$\vec{B}_{SOC}(q_{i}, {\bf n}_{{\bf r}_i},\Omega,\delta)=\left[0,\Omega, \Delta(q_{i}, {\bf n}_{{\bf r}_i})-\delta\right]$ is an effective $q$-dependent magnetic field arising from the SOC term $\Delta(q_{i}, {\bf n}_{{\bf r}_i})=(E\left( q_{i},{\bf  n}_{{\bf r}_i}\right ) -E\left (q_{i}+\phi, {\bf  n}_{{\bf r}_i}\right))/\hbar$. The operators  ${\bf \hat S}_{i}$ are spin-$1/2$ angular momentum operators acting on the two clock states of atom $i$.

Figure~\ref{fig:Cartoon}(a) displays the spin-orbit coupled bands. In the tight-binding approximation the largest detunings from the bare transition frequency are given by $\delta_{\pm}=\Delta(q\sim\{0,\pi\}, {\bf{n_{r}}})=\pm4J\sin\left(\phi/2\right)$. Here, $J$ is the thermally averaged value of the tunneling rate (see Methods). When $\Delta(q, {\bf{n_{r}}})>\Omega$, the SOC broadening of the lineshape is spectroscopically resolved and exhibits two peaks at clock laser detunings of $\delta_\pm$ (Fig.~\ref{fig:Cartoon} (b)).  These peaks arise from divergences of the joint density of states called Van Hove singularities\cite{Kolkowitz2017}.

To observe the dynamics of our spin-orbit coupled system we perform Ramsey spectroscopy. A strong initial pulse of area $\theta_{1}$ and $\delta=0$ excites all atoms into a coherent  superposition of clock states that are then allowed to freely evolve during $\tau$.  Although the clock laser is off during this period, the atoms accumulate phase in the rotating frame of the laser and thereby retain the imprinted optical phase. As a result, the atoms continue to experience the SOC induced effective magnetic field $\vec{B}_{SOC}(q_{i}, {\bf n}_{{\bf r}_i}$, $\Omega$=0, $\delta$=0) throughout the dark time $\tau$. One observable we measure using this procedure is the Ramsey fringe contrast, $\mathcal{C}=2\sqrt{\left\langle\hat{S}^{x}\right\rangle^{2}+\left\langle\hat{S}^{y}\right\rangle^{2}}/N$, which is the length of the projection of the collective magnetization in the $\hat{e}_x-\hat{e}_y$ plane of the Bloch sphere.
Here  ${\bf \hat S}=\left[\hat{S}^{x},\hat{S}^{y},\hat S^{z}\right]$ are collective spin operators with ${\hat S}^{\{x,y,z\}}=\sum_{i=1}^{N}{\hat S}_{i}^{\{x,y,z\}}$. 

The concentration of atoms at the two Van Hove singularities allows us to qualitatively understand the ensuing dynamics as arising mainly from these two groups of atoms, with quasimomenta $q\sim0$ and $q\sim\pi$, and corresponding detunings of $\delta_{\pm}$, respectively. Figure~\ref{fig:Cartoon}(c) depicts the Bloch sphere visualization of Ramsey spectroscopy for the case when the two groups are non-interacting and for $\theta_{1}=\pi/2$. For a variable evolution time $\tau$, the atoms with opposite detunings $\delta_{\pm}$ evolve around the equator of the Bloch sphere in opposite directions (dashed blue arrows). Consequently, the length of the collective spin vector (solid blue arrow) changes, but the vector direction remains parallel or anti-parallel to $\hat{e}_{x}$.

Representative single-particle contrast curves are shown in Fig.~\ref{fig:Cartoon}(d) for tunneling rates $J_{1}/(2\pi)=3.2$ Hz (green triangles) and $J_{2}/(2\pi)=17.6$ Hz (blue circles) as a function of $\tau$.  
  This data was taken in the non-interacting regime by using a small number of atoms ($N<500$). The collapses and revivals in the contrast can be readily understood from the simple model of the two atom groups. When the two groups of atoms accumulate a phase difference of $\pi$ the length of the collective Bloch vector will be zero ($\mathcal{C}=$ 0).  The detuning, $\delta_{\pm}\approx\pm4J$ determines the precession rate around the Bloch sphere and we thus expect the contrast to collapse and revive with a periodicity proportional to $1/J$.  In Fig.~\ref{fig:Cartoon}(e) the x-axis is scaled as a function of $J\tau/(2\pi)$, illustrating that the contrast curves for different $J$ values then collapse onto a single curve.


An obvious feature of the observed contrast evolution is the long term decay, which is not captured by the simple two-group approximation. While the joint density of states is the largest at the Van Hove Singularities, all $q$ values are in fact populated, with atoms contributing at detunings in-between $\delta_{\pm}$. Summing over the contributions from all $q$, the resulting time-dependence of the contrast is given by $\mathcal{C}=\sin(\theta_{1})|\mathcal{J}_{0}\left[4J\tau \sin\left(\phi/2\right)\right]|$, where ${\mathcal J}_{0}$ is a zeroth order Bessel function of the first kind, as shown in Fig.~\ref{fig:Cartoon}(d,e).

To validate that the collapses, revivals, and overall decay in contrast are due to $B_{SOC}(q)$, we can remove its effect by adding a spin echo pulse to the Ramsey sequence. Any dephasing from the static SOC-induced effective magnetic field during the first $\tau/2$ period of free evolution will re-phase during the second $\tau/2$ free evolution period, due to the $\pi$ echo pulse, which flips the sign of phase accumulation.  Figure~\ref{fig:Echo}(a) shows the effect of spin echo (orange diamonds) for $J/(2\pi)=4.2$ Hz.  The spin echo eliminates the collapses and revivals from the Ramsey fringe contrast (purple circles), and prolongs the overall contrast decay. However, the observed decay in contrast at long times is still fast compared to the contrast decay time for $J=0$ ($\sim1$ s, see Methods). Contrast decays under spin echo for different values of $J$ are shown in Fig.~\ref{fig:Echo}(b). The spin echo decays do not collapse to a single curve when the free evolution time is scaled to $J\tau/(2\pi)$, indicating that the additional dephasing does not scale linearly with $J$, and that quasimomentum is not conserved at large $\tau$. It is this additional dephasing that results in the suppression of the revivals in contrast at lower tunneling rates as shown in Fig.~\ref{fig:Cartoon}(d,e) (green triangles).  Throughout the rest of this work we incorporate the empirically observed $J$ dependence of this dephasing, which we call diffusive dephasing, into our model (see Methods).


Having characterized the single particle dynamics under SOC, we introduce interactions by increasing the atomic density. As shown in Fig.~\ref{fig:Contrast}, signatures of strong spin interactions start to emerge as $N$ increases. The blue circles are the case of no interactions ($N<$ 500) and red squares are the case where we introduce interactions by increasing the atom number by more than an order of magnitude ($N\sim1\times10^{4}$).  We observe in Fig.~\ref{fig:Contrast} (a) that for an initial Ramsey pulse $\theta_{1}=\pi/4$, an increase in atomic density qualitatively alters the dynamics, suppressing the collapses in contrast observed for the low density case. For $\theta_{1}=\pi/2$ we observe that interactions shift the zeroes of the contrast compared to the non-interacting case (Fig.~\ref{fig:Contrast} (c)), and by further increasing the density and reducing the tunneling rate, we see that the first collapse of the contrast can be suppressed altogether (Fig.~\ref{fig:Contrast} (d)).  We note that for $J=0$ the contrast decay has been previously seen to be highly sensitive to the initial Ramsey pulse area and no contrast revival with interactions has been observed\cite{MJMManyBodySr,XiboSUN}.   

 In order to quantitatively understand the complex interplay between interactions and SOC, we consider the spin model that has previously been successfully used to understand many-body interactions in optical lattice clocks\cite{MJMManyBodySr,XiboSUN}. During these measurements all atoms are initially prepared in the $\ket{g}$ spin state, and each atom occupies a single motional mode in the lattice. The initial mode distribution is preserved during clock interrogation as the collision energy is insufficient to alter the motional eigenstates. Since the motional degrees of freedom are frozen, we can treat the single-particle modes as corresponding lattice sites spanning an energy space. Thus, $s$-wave and $p$-wave contact interactions are  mapped into non-local, infinite-range collective interactions between the electronic pseudospins in the energy-space lattice\cite{MJMManyBodySr}. The Hamiltonian for our system, including interactions, then becomes $\hat{H}=\hat{H}_{SOC} +\hat{H}_{\rm int}$, with $\hat{H}_{\rm int}$ given by,
\begin{equation}
\label{eq:IntHamiltonian}
	{\hat H}_{\rm int}/\hbar=\frac{\chi}{L}\left({\hat S}^{z}\right)^{2}+\frac{C}{L}\left(N\right){\hat S}^{z}+\frac{\xi}{L} {\bf \hat S}\cdot {\bf \hat S}.
\end{equation}
\noindent  The spin couplings $\chi=\left(V_{gg}+V_{ee}-2V_{eg}\right)/2$, $C=\left( V_{ee}-V_{gg}\right)/2$, and $\xi=\left(V_{eg}-U_{eg}^-\right)/2$ depend on  $V_{\alpha\beta}$ and $U_{\alpha\beta}^-$ which are the $p$-wave and $s$-wave mean interaction parameters, respectively. $L$ is the number of lattice sites, and thus $N/L$ represents the mean number of atoms per site.  Due to the temperature of the atoms in the lattice being $>1\mu$K the $s$-wave and $p$-wave interactions are similar in magnitude (see Methods).

The term proportional to $\xi$ encapsulates the exchange interaction process mediated by both ${s}$-wave and $p$-wave collisions.  For the nuclear spin-polarized identical fermions initially prepared in the lattice, and in the absence of SOC, this term becomes a constant of motion and is thus irrelevant to the dynamics. 
However, when $J\neq$ 0, the effective $q$-dependent SOC magnetic field $\vec{B}_{SOC}(q_{i}, {\bf n}_{{\bf r}_i},0,\delta)$  causes the initially spin polarized atoms to dephase with respect to each other, thereby introducing exchange interactions between them, which directly compete with the single-particle SOC dynamics.

The $p$-wave interaction terms proportional to $\chi$ and $C$ generate a collective Ising Hamiltonian which commutes with ${\bf {\hat S}}^2$ and have previously, in the absence of SOC, been shown to induce  many-body spin dynamics\cite{MJMManyBodySr}  for any superposition of $e$ and $g$. These terms are unchanged in the presence of SOC, and have a negligible effect on the spin contrast for the experimental conditions and timescales we study here .


Throughout this work, we find that the explored experimental timescales are in a regime where the mean field approximation is valid.  In this approximation the interaction terms can be treated as an additional time-dependent magnetic field generated by the collective spin vector, $\vec{B}_{\rm int}$. This allows us to factor out a collective spin operator from $\hat{H}_{\rm int}$ as given in Eq.~\ref{eq:IntHamiltonian} in order to arrive at the mean field Hamiltonian including both interactions and SOC:
\begin{equation}
\label{eq:MFHamilton}
	\hat{H}^{\rm MF}/\hbar=-\sum_{i=1}^N \vec{B}_{SOC}(q_{i}, {\bf n}_{{\bf r}_i},\Omega,\delta) \cdot {\bf \hat S}_{i}+\sum_{i=1}^N \vec{B}_{\rm int}\cdot {\bf \hat S}_{i}
\end{equation}
\noindent where $\vec{B}_{\rm int}=\left [ \frac{2\xi}{L} \left\langle\hat{S}^{x}\right\rangle, \frac{2\xi}{L} \left\langle\hat{S}^{y}\right\rangle ,  \left (2\frac{\xi+\chi}{L}\right ) \left\langle\hat{S}^{z}\right\rangle +N \frac{C}{L}\right]$.  The $\hat{e}_{x}$ and $\hat{e}_{y}$ components can be written together as a collective, evolving, transverse magnetic field, $\frac{ 2\xi}{L}\langle {\bf {\hat  S}}^{\bot}(t)\rangle$ around which individual atom Bloch vectors rotate.  This term competes with the single particle dephasing term, $\vec{B}_{SOC}(q_{i}, {\bf n}_{{\bf r}_i},\Omega,\delta)$, and forces the pseudo-spins to remain aligned, causing interaction dependent changes to the contrast.  The $\hat{e}_{z}$ component of the interaction magnetic field is a constant of motion and gives rise to a collective precession of the Bloch vectors at a rate $N\Big (\frac{C}{L}- \frac{\chi+\xi}{L}\cos\theta_1 \Big){\bf }$, where $\langle\hat{S}^{z}\rangle=-N/2 \cos\theta_{1}$.  When the tunneling rate $J$ is zero, all the terms proportional to $\xi$ in $\hat{H}^{MF}$ will not affect the contrast or frequency shift.
The competition between the interaction-induced transverse magnetic field and the static SOC dephasing is shown schematically in Fig.~\ref{fig:Contrast}(b) and \ref{fig:Contrast}(e) under the simple two-group approximation for $\theta_{1}=\pi/4$ and $\theta_{1}=\pi/2$ respectively.  For $\theta_{1}=\pi/4$, the collective rotation differentially changes the projected length of the individual Bloch vectors on the transverse plane, generating a net $|\langle {\hat S}^{y}(t)\rangle|>0 $.  As a result, when the vectors are $\pi$ out of phase, they no longer completely cancel, leaving a finite contrast at all times, as opposed to the complete collapse observed for the non-interacting case  where $|\langle {\hat S}^{y}(t)\rangle|=0 $.  This is apparent in the data shown in Fig.~\ref{fig:Contrast}(a), where  the contrast remains finite  for the interacting case (red circles, $N\xi/L=-2.0$ Hz, $N\chi/L=1.2$ Hz).

For $\theta_{1}=\pi/2$, due to symmetry, the rotation of the Bloch vectors (red, dashed arrows in Fig.~\ref{fig:Contrast}(e)) around the collective spin vector (red, solid arrow) does not change the relative transverse length of the vectors -- which imposes $|\langle {\hat S}^{y}(t)\rangle|=0 $. The effects of interactions are shown in Fig.~\ref{fig:Contrast}(c)-(d) for varying strengths of interactions ($N\xi/L$) compared to $J$.  When the interactions are still small compared to the tunneling ($J>\lvert N\xi/L\rvert$, with $N\xi/L=-3.5$ Hz and $N\chi/L=1.3$ Hz) (Fig.~\ref{fig:Contrast}(c)), they cause no qualitative change to $\mathcal{C}$ compared to the non-interacting case, except for a weak rephasing of the spins that slightly delays the contrast collapse and decreases the revival amplitude. This is manifested as an interaction-induced shift of the time of the first contrast zero, $\propto N^2\xi^{2}/L^2 J^{2}$.

If $J$ is decreased such that $J\sim \lvert N\xi/L\rvert$, then the exchange interactions produce a qualitatively different behavior, as shown in Fig.~\ref{fig:Contrast}(d). For $J/\left (2\pi\right )=1.3$ Hz, the non-interacting case (blue circles) shows the characteristic collapse and revival. In contrast, the interacting case (with $N\xi/L=-5.6$ Hz, $N\chi/L=3.4$ Hz), shows no collapse whatsoever, instead exhibiting only a monotonic decay with $J\tau$. The suppression of the collapse and revivals is a result of the exchange-induced rephasing of the spins (Fig.~\ref{fig:Contrast}(e)). Ideally, this type of spin locking would preserve the coherence indefinitely, as can been seen directly from the interacting Hamiltonian (\ref{eq:IntHamiltonian}), where for large $\xi$  the initial state is an eigenstate. Indeed, long-term synchronization has been previously observed in other cold atom experiments\cite{LongClockCoherence} with dominant $s$-wave interactions. In our OLC we also need to account for competing mechanisms.

One important decoherence mechanism is atom loss due to inelastic two-body $e-e$ $p$-wave collisions\cite{ManyBodyTheory,BishofInelastic}, which becomes particularly relevant for a large $N$. The effect of the losses on the contrast, however, is largely compensated when the contrast is normalized by the total atom number, as we do throughout this work (see Methods).
The most relevant contribution to decoherence for the current experiment is the single-particle diffusive dephasing observed in Fig.~\ref{fig:Echo}. Its effect on the contrast can already be seen in the non-interacting case (blue circles) and is exacerbated when operating at the low tunneling rates required to enter the $J\sim \lvert N\xi/L\rvert$ regime. 
We anticipate that quasimomentum conservation, and signatures of spin-locking at longer times,  will be achievable in a 3D optical lattice, where  coupling to the thermally populated radial modes would be eliminated (see Methods). 

To complete our full characterization of the spin system and to disentangle the interaction dynamics from decoherence, we also study the effects of interactions on the phase accumulated by the collective spin vector during the free precession time $\tau$,
	$\tan (\Delta \nu 2 \pi \tau)= \left\langle\hat{S}^{y}\right\rangle / \left\langle\hat{S}^{x}\right\rangle$.  
In optical lattice clocks this is traditionally described by a density-dependent frequency shift\cite{MJMManyBodySr, LemkePwave} ($\Delta \nu$).  

 For $J=0$ (no SOC), $\xi$ is a constant of motion, and the density shift arises entirely from the Ising $p$-wave interactions. In Fig. \ref{fig:Shift}(a) a density shift measurement done without SOC  shows a  clear linear dependence  on the  fraction of the atoms in the excited clock state, fully consistent with  previous works\cite{MJMManyBodySr,XiboSUN}. There for $J=0$,  $\Delta\nu$ has been well characterized and found to be independent of the dark time between the Ramsey pulses. In this work the measured shift in the absence of SOC agrees with the prediction from the mean-field Hamiltonian (Eq.~\ref{eq:MFHamilton}) $\Delta\nu_{J=0}=N\left (\frac{C}{L} -\frac{\chi}{L}\cos{\theta_{1}}\right )$, where the shift depends linearly on the fraction of atoms in the excited clock state ($P_{e}=\left(1-\cos{\theta_{1}}\right)/2$).

In contrast, interactions in the presence of SOC give rise to a frequency shift that is dependent on the dark time between the Ramsey pulses (Fig. \ref{fig:Shift}(c,d)). From this data the frequency shift extrapolated to zero excitation fraction is seen to diverge when the single-particle contrast  decays to zero (see Fig. \ref{fig:Shift}(d)).

For $J>0$  (with SOC),  the situation becomes more complicated. To develop an intuitive understanding, we return again to the two atom-group model, where a simple analytic expression can be derived to first order in interactions, (for a more generic lattice treatment, see Methods) 
\begin{equation}
\label{eq:SPShift}
	\Delta\nu=\Delta\nu_{J=0}-\frac{N \xi}{L}\cos(\theta_{1})\left (1-\frac{\tan(4J\tau)}{4 J\tau}\right ).
\end{equation} The same exchange term that produces the time-dependent collective transverse field responsible for modifying the contrast dynamics also results in a frequency shift. This term diverges when $\cos(4J\tau)=0$, which physically corresponds to the case when the two non-interacting atom group vectors are $\pi$ radians out of phase on the Bloch sphere, as illustrated in Fig.~\ref{fig:Shift}(e). When the spins rephase and the contrast becomes finite again, the exchange-induced shift diminishes. It completely turns off in the two atom-group approximation when the spins re-align. However, for the experimentally relevant case of a thermally populated band with all $q$ values participating, the density shift will change in magnitude with time but will not disappear completely, since the spins do not completely rephase.

The experimentally measured dependence of the SOC density shift on $\tau$ at a finite tunneling rate of $J/\left(2\pi\right)=2.2$ Hz is shown in  Fig.~\ref{fig:Shift}(c)-(d) for $N\xi/L=-2.7$ Hz and $N\chi/L=1.6$ Hz. The observed shift is not entirely linear in excitation fraction, indicating that the interactions can no longer be described by first order perturbation theory, and higher order corrections are required (see Methods).  Figure~\ref{fig:Shift}(d) compares the contrast to the extrapolated density shift for zero excitation fraction ($\widetilde{P}=P_{e}\Big\rvert_{e=0}$) for the same data as in Fig.~\ref{fig:Shift}(c).  The extracted quantity $\widetilde{P}$ shows a divergence around the zero of the contrast, consistent with Eq.~\ref{eq:SPShift}. The highly non-trivial functional form of the density shift indicates that SOC induced exchange interactions will be a major factor in optical lattice clocks if the effects of tunneling are not suppressed. 
However, the experimentally observed density shift and contrast, which encapsulate the magnetisation dynamics, can be well described by theory.  This agreement highlights the fact that for the experimentally relevant timescales, the complex interplay between SOC and many-body dynamics can be understood and explored precisely.


In conclusion, we have explored the emergence of complex dynamics with interacting fermions under engineered spin-orbit coupling in a Sr optical lattice clock. The many-body dynamics are fully characterized by a collective XXZ Hamiltonian aside from extra dephasing arising from non-conserved quasimomenta.
In the future we plan to suppress this dephasing by using more sophisticated pulse sequences\cite{DynamicalDecoupling} or by employing a 3D optical lattice, where the $p$- wave interactions would also be suppressed. The lower temperatures associated with loading a Fermi-degenerate gas in a 3D lattice geometry\cite{Campbell} will also enable the study of SOC in higher dimensions, precise control of the SOC phase\cite{FallaniNew,GreinerSOC2016} $\phi$, and exploration of a new strongly interacting regime where the collective XXZ model is no longer applicable, and where richer exotic behaviors including topological superfluids\cite{Qi2011} and Kondo correlated metallic phases can emerge\cite{Isaev2016}.

\noindent\textbf{Acknowledgements}\\
We are grateful to M. Lukin, S. Yelin, V. Gurarie, M. Foster, S. L. Campbell, A. Goban, R. B. Hutson, G.E. Marti, E. Oelker, J. Robinson, L. Sonderhouse,  and D. X. Reed for stimulating discussions and technical contributions.  We thank M. Norcia and A. Kaufman for their careful reading of the manuscript.  This research is supported by NIST, DARPA, JILA Physics Frontier Center (NSF-PFC-1125844), AFOSR-MURI, and AFOSR. C.S. is partially supported by the JILA Visiting Fellow Program.

\noindent\textbf{Author contributions}\\
S.L.B., S.K., T.B., D.K., and J.Y. contributed to the executions of the experiments. A.S.-N., M.L.W. and A.M.R. developed the theory model. All authors discussed the results, contributed to the data analysis and worked together on the manuscript.

\bibliographystyle{naturemag}

\newpage

\begin{figure*}

  \centering
  \includegraphics[scale=0.48]{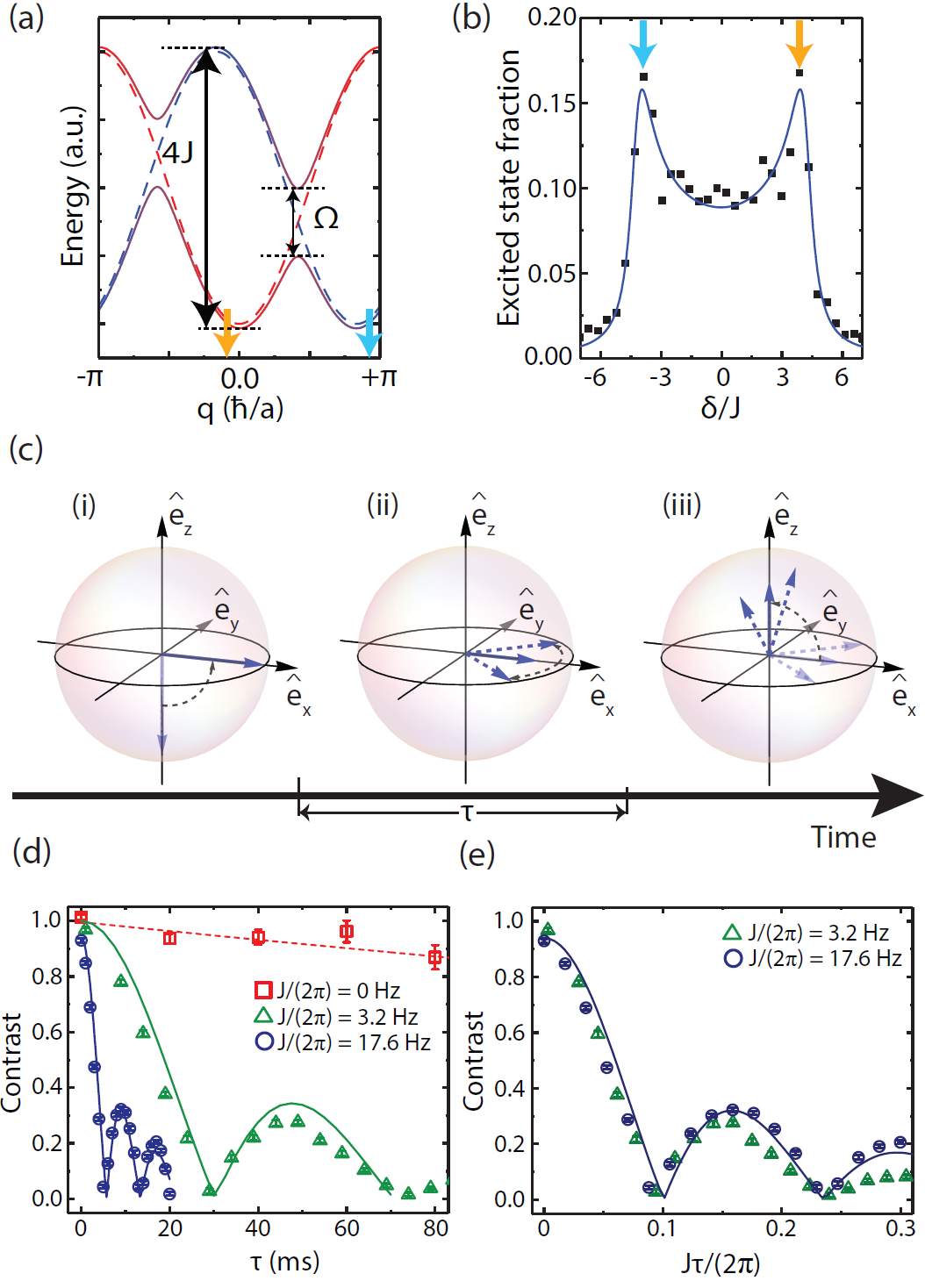}
  \caption[The short caption]{\textbf{Ramsey spectroscopy with spin-orbit coupling. (a)}  Spin-orbit coupled bands (solid lines) due to the coupling of the ground bands of the bare clock state $\ket{g, q}_{\bf{n_{r}}}$ (red, dashed) and the momentum-shifted clock state $\ket{e, q+\phi}_{\bf{n_{r}}}$ (blue, dashed) with a band splitting given by the Rabi frequency, $\Omega$, and a bandwidth $4J$, where $J$ is the tunneling rate and $\phi=\pi\lambda_{L}/\lambda_{c}$.  Van Hove singularities (VHSs) occur at quasimomenta $q\sim0$ and $q\sim\pi$ (yellow and blue arrows).  \textbf{(b)} SOC results in a split atomic lineshape with the VHSs at clock laser detunings $\delta_{\pm}=\pm 4J\sin(\phi/2)$ (yellow and blue arrows).  Data is black squares and theory fit is solid line.  \textbf{(c)} Bloch spheres for single particle Ramsey dynamics. The two subsets of atoms at the two VHSs, with $\delta_{\pm}$, approximate our system. (i) Both atom groups initially start in $\ket{g, q}_{\bf{n_{r}}}$. A strong pulse of area $\theta_{1}=\pi/2$ ($\delta=0$) rotates them around $\hat{e}_{y}$ to a superposition of $\ket{g, q}_{\bf{n_{r}}}$ and $\ket{e, q+\phi}_{\bf{n_{r}}}$.  (ii) During $\tau$ the VHSs precess around the Bloch sphere in opposite directions changing the length of the collective spin vector (solid arrow).  (iii)  A readout pulse of area $\theta_{2}=\pi/2$ extracts the Ramsey fringe contrast by measuring the excited state fraction.  (Continued on the following page.)}
\end{figure*}
\begin{figure*}
  \contcaption{\textbf{(d)} The contrast decay for different tunneling rates $J_{1}/(2\pi)=3.2$ Hz (green triangles), $J_{2}/(2\pi)=17.6$ Hz (blue circles), and $J_{3}/(2\pi)=0$ Hz (red circles) as a function of $\tau$ without interactions.  \textbf{(e)}  The same data as in \textbf{(d)}, with the $x$-axis scaled to $J\tau/(2\pi)$. The solid lines are theory curves and the dashed line is an exponential fit with a decay constant of $\sim 0.6$ s, and the error bars are $1\sigma$ confidence intervals.}
\label{fig:Cartoon}
\end{figure*}

\newpage
\clearpage
%
\begin{figure*}
  \centering
  \includegraphics[scale=0.5]{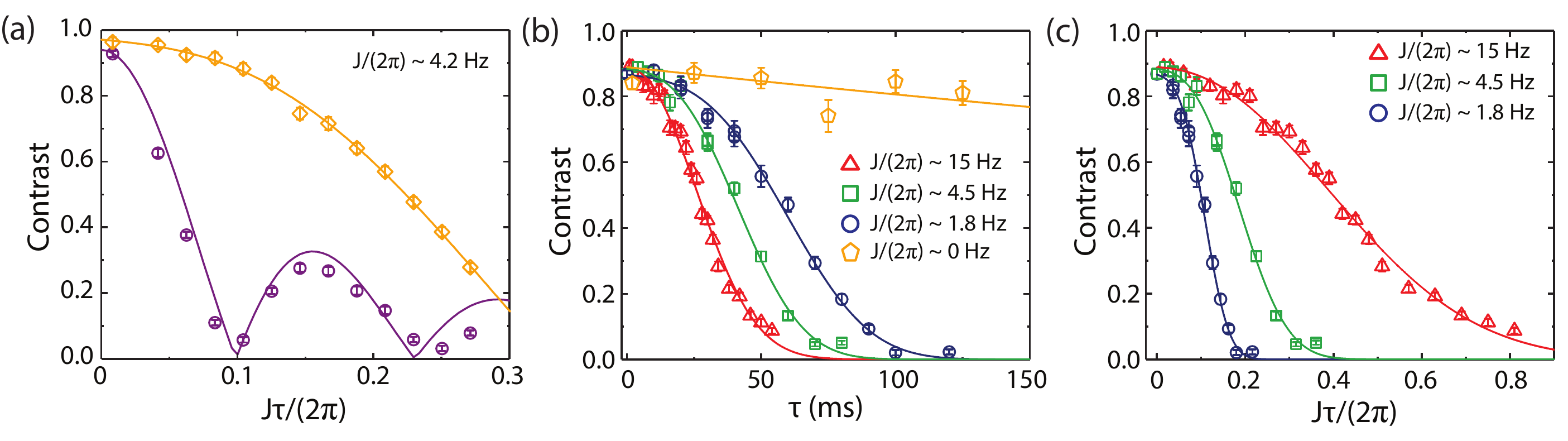}
  \caption[The short caption]{\textbf{Non-interacting echo decay. (a)} A spin echo pulse (orange, diamonds) removes the static dephasing caused by SOC for a Ramsey sequence for the same $J$ (purple circles, theory is solid purple line.) \textbf{(b)}  Spin echo decay of contrast for four different $J$ values  as a function of $\tau$.  \textbf{(c)}  Same data as in  \textbf{(b)} (for non-zero $J$) as a function of $J\tau$. All $J>0$ curves decay as $\propto e^{-\left(\tau/\tau_{d}\right )^{3}}$, where $\tau_{d}$ is a nonlinear function of $J$, implying an extra diffusive dephasing, and do not collapse to a single curve when the free precession time is rescaled to $J\tau$ (See Methods).  For $J=0$ the echo data decays $\propto e^{-\left(\tau/\tau_{d_0}\right )}$.  All error bars are from individual contrast fits, and all solid lines for the $J>0$ ($J=0$) echo data are fits $\propto e^{-\left(\tau/\tau_{d}\right )^{3}}$ ($\propto e^{-\left(\tau/\tau_{d_0}\right )}$). 
 }
\label{fig:Echo}
\end{figure*}

\newpage
\clearpage
\begin{figure*}

  \centering
  \includegraphics[scale=0.6]{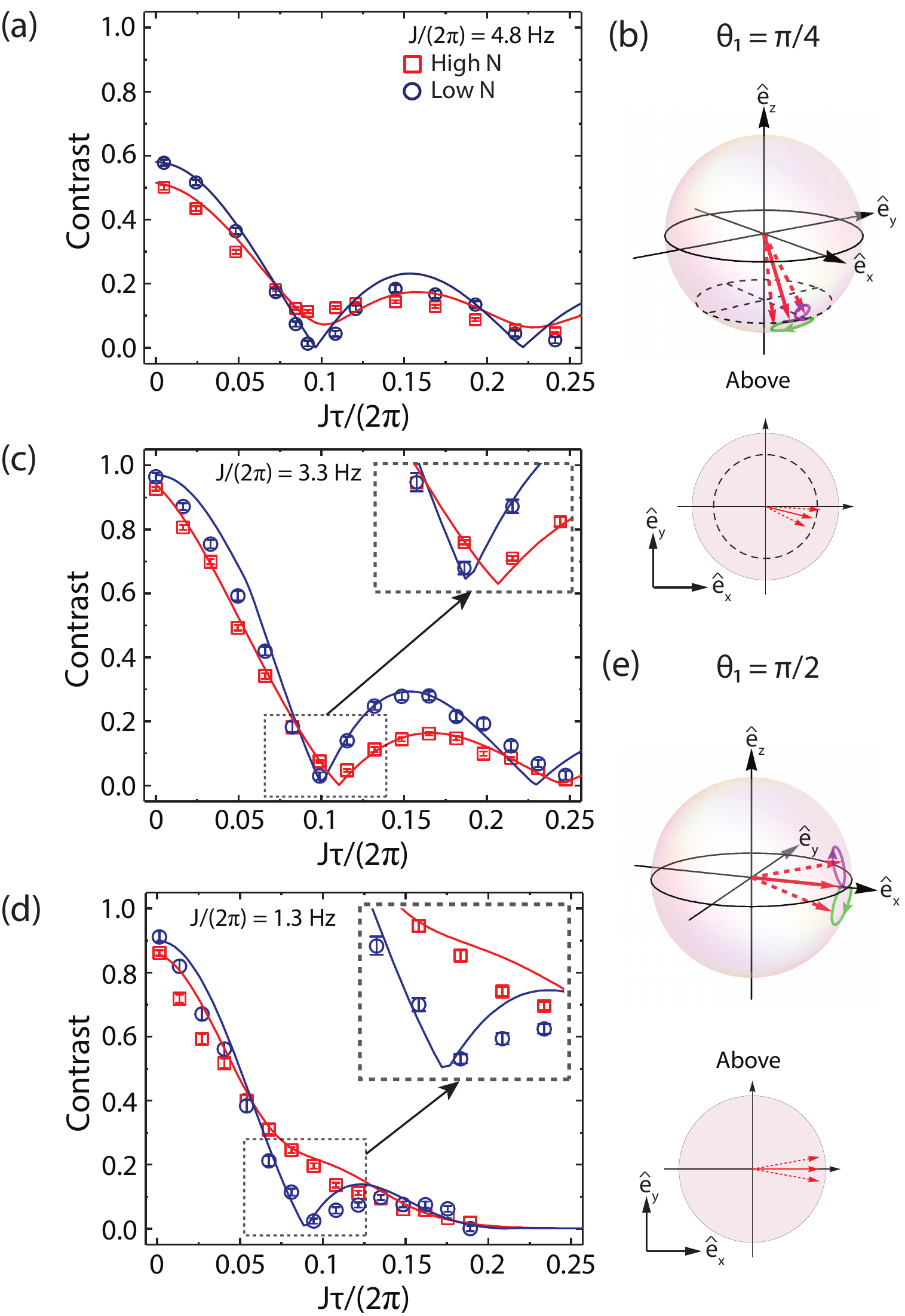}
\caption{(Continued on the following page.)}
\label{fig:Contrast}
\end{figure*}
\begin{figure*}
  \contcaption{\textbf{Spin-orbit coupling with varying interactions. (a)}  Contrast data for high density (red squares) and low density (blue circles) for $\theta_{1}=\pi/4$.  The high-density contrast does not go to zero due to the rotation caused by exchange interactions.  \textbf{(b)}  As the VHSs (dashed arrows) rotate around the Bloch sphere they become distinguishable and exchange interactions induce rotations shown as purple and green trajectories.  For $\theta_{1}=\pi/4$  this rotation leads to the Bloch vectors for the two VHSs being of unequal length in the $\hat{e}_{x}-\hat{e}_{y}$ plane so that the collective spin vector (solid arrow) remains finite.  \textbf{(c)} and \textbf{(d)} are contrast curves for $\theta_{1}=\pi/2$.  \textbf{(c)} $J>\lvert N\xi/L= -3.5$ Hz$\rvert$ and the interactions cause the zero in the contrast to be pushed to larger $J\tau$.  \textbf{(d)}  $J\sim \lvert N\xi/L=-5.6$ Hz$\rvert$ and interactions prevent static dephasing and the contrast approaches zero only at long times.  Solid lines are theory including atom loss and diffusive dephasing (see Fig.~\ref{fig:Echo}).  \textbf{(e)}  For $\theta_{1}=\pi/2$ (bottom) the exchange induced rotation is symmetric and the two Bloch vectors are the same length in the $\hat{e}_{x}-\hat{e}_{y}$ plane.}
\label{fig:Contrast}
\end{figure*}
\newpage
\clearpage

\begin{figure*}
  \centering
  \includegraphics[scale=0.6]{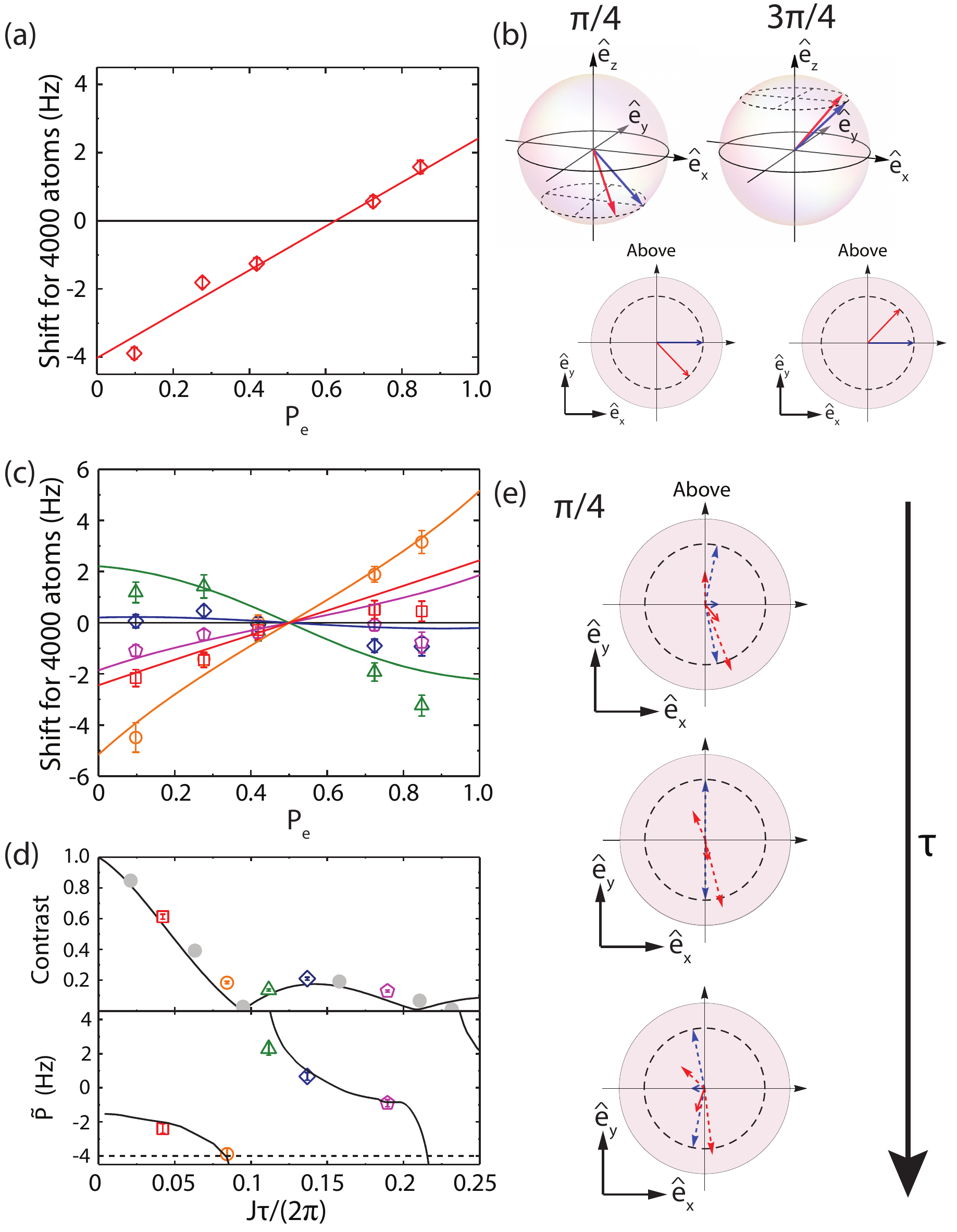}
  \caption{(Continued on the following page.)}
  \label{fig:Shift}
\end{figure*}
\begin{figure*}
  \contcaption{\textbf{Interactions with and without spin-orbit coupling.  (a)} With $J=0$ (no SOC) atoms interact via $p$-wave collisions only, leading to a frequency shift linearly dependent on excitation fraction, $P_{e}$. \textbf{(b)} On the Bloch sphere for $\theta_{1}<\pi/2$ the collective vector for atoms with interactions (red) rotates to give a negative phase shift with respect to the non-interacting (blue) vector.  For $\theta_{1}>\pi/2$ the interactions lead to a positive shift.  \textbf{(c)}  Density shifts for $J/(2\pi)=2.2$ Hz (with SOC) for different $J\tau$ measured by varying $\tau$, as indicated in panel \textbf{(d)}.  The magnitude and sign of the density shift can be seen to vary with time.  \textbf{(d)}  The non-interacting contrast curve,  including additional grey contrast data, and the $\widetilde{P}=P_{e}\Big\rvert_{e=0}$ density shift for 4000 atoms corresponding to the data in \textbf{(c)} including theory curves for $J>0$ (solid, black) and $J=0$ (dashed, black).  \textbf{(e)}  The divergence can be understood by considering the Bloch vectors for the different VHSs (dashed arrows) with $\tau$.  As the contrast goes through zero for no interactions (blue), the SOC induced exchange interactions (red) prevent the collective spin vector (red, solid arrow) from going to zero.  As the non-interacting collective spin vector (blue, solid) goes through zero it changes sign, causing a change in sign of the slope of the density shift.  }
\end{figure*}
\newpage
\clearpage

\section{Experimental Methods}
In our experiment the lattice is formed by a $\sim3$ W incoming laser beam that is focused to a waist of $\omega_{0}=40 \mu$m at the position of the atoms. After exiting the vacuum chamber, the laser beam is collimated and then retro-reflected back on itself after passing through two acousto-optical modulators (AOMs). These AOMs are used to dynamically ramp the lattice depth without changing the frequency of the retro-reflected beam.  During each experimental cycle the atoms are loaded into a lattice of depth $U_{z}/E_{R}=130$, where $E_{R}$ is the recoil energy.  This lattice depth corresponds to an axial trapping frequency of $\nu_{z}\approx2E_{R}\sqrt{U_{z}/E_{R}}/(2\pi\hbar)=80$ kHz.  The atoms are then sideband cooled to the ground band, and then the lattice is ramped down adiabatically to give the desired tunneling rate.  To measure interaction effects the atom number is varied without changing the initial distribution of atoms within the lattice.

To measure the contrast following a Ramsey or spin echo sequence, the laser detuning is kept constant and the phase of the second $\pi/2$ Ramsey pulse is varied with respect to the phase of the first $\pi/2$ Ramsey pulse during many cycles of the experiment to produce a Ramsey fringe.  The contrast is extracted by fitting the Ramsey fringe with a sinusoid, with phase and amplitude as the only free parameters. To measure the contrast change with density,  Ramsey fringe contrast measurements are taken for high and low atom densities at each dark time.  Each Ramsey fringe measurement consists of $\sim 80$ individual experimental cycles where the cycle time is $\sim1.5$ s.

For the frequency shift measurement at $J/(2\pi)\sim 0$ Hz, for different initial pulse areas and a set $\tau=80$ ms free precession time, the phase of the final $\pi/2$ Ramsey pulse is again varied while interleaving measurements with the atom number switching between $\sim 300$-$4000$ atoms. The Ramsey fringes are then again fitted with sinusoids and the phase difference between the high density and low density case is extracted and converted to a frequency shift.  Between each data set the excitation fraction is also measured by applying only the first Ramsey pulse, then measuring the clock state populations.  For $J\neq 0$ the same process is repeated for different dark times. To extract $\widetilde{P}$ we fit each experimental density shift measurement with a linear fit.

\begin{figure*}
  \centering
  \includegraphics[scale=0.5]{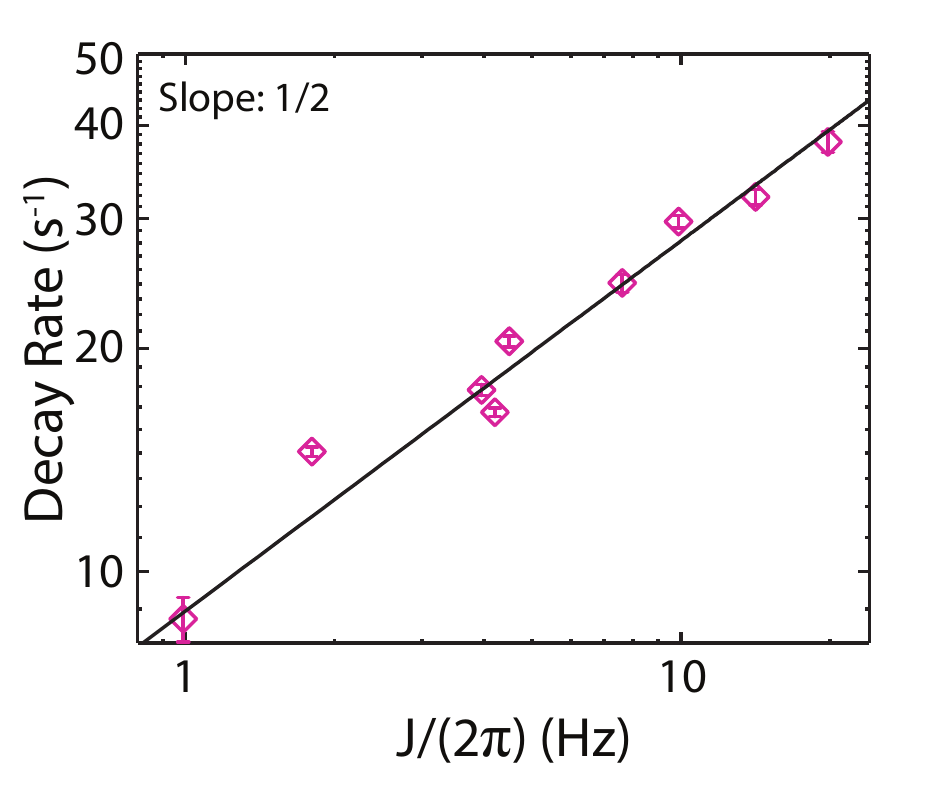}
 \caption*{\textbf{Extended Data Figure 1: Diffusive dephasing}  The decay rates ($1/\tau_{d}$) as a function of $\log J$ for different lattice depths show a $J$ dependence consistent with a $1/\tau_{d}\propto J^{1/2}$. The error bars come from fits of $\tau_{d}$ and the solid line is a best fit curve using $1/\tau_{d}\propto J^{1/2}$.}
  \label{fig:ED1}
\end{figure*}

For $J>1$ Hz, the SOC-induced dephasing dominates over all other dephasing mechanisms in our clock. While our Hamiltonian accounts for the static dephasing, a systematic investigation of the relevant range of $J$ reveals that the spin echo data has an additional decay with the functional form of $\propto e^{-\left(\tau/\tau_{d}\right )^{3}}$ as shown in Fig.~\ref{fig:Echo} (b) in the main text. This form of spin echo decay is well known from NMR and solid-state spin defect experiments\cite{Lange, Slichter}, where the dephasing is the result of a slow, random diffusion of magnetic field with time. In our case this corresponds to a diffusion of the SOC effective magnetic field $B_{SOC}(q)$, indicating that the quasimomentum is not conserved at large $\tau$. The extracted decay rate $\left(1/\tau_{d}\right)$ for different $J$ is shown in Extended Data Fig.~1. The scaling is consistent with $1/\tau_{d}\propto J^{1/2}$. We observe the same scaling of decay rate with $J$ for the Ramsey sequences, and include this decay in our theory model. The most likely mechanism for this empirically observed diffusion of $B_{SOC}$ is the coupling of axial motion to the thermally populated radial modes. The spatial inhomogeneity in $J$ across the lattice due to the finite Rayleigh range of the lattice beams may also contribute.

\section{Theoretical Methods}

\subsection{ Hamiltonian}

The Hamiltonian governing a nuclear spin-polarized ensemble of fermionic atoms   with two accessible clock levels, $^1S_0 (g)$ -${}^3P_0 (e) $, which are controlled by a linearly polarized clock laser beam\cite{ManyBodyTheory}  is given in the rotating frame of the clock laser by:
{\small\begin{eqnarray}
\label{eq:H} &&\hat{H}=\hat{H}_0+\hat{H}_L+\hat{H}_{int},\\
&&\hat{H}_0=\sum_{\alpha} \int d^3\mathbf{R}\,\hat{\psi}_{\alpha}^\dagger\left(\mathbf{R}\right)\left[-\frac{\hbar^2}{2m}\nabla^2+V_{\mathrm{ext}}\left(\mathbf{R}\right)\right]\hat{\psi}_{\alpha}\left(\mathbf{R}\right),\\
&&\hat{H}_L=-\frac{\hbar \Omega}{2}\int d^3\mathbf{R} \left[\hat{\psi}_{e}^{\dagger}\left(\mathbf{R}\right)e^{i2\pi Z/\lambda_c}\hat{\psi}_{g}\left(\mathbf{R}\right)+\mathrm{H.c.}\right] -\frac{ \hbar  \delta}{2}  \int d^3\mathbf{R}\left[\hat{\psi}_{e}^{\dagger}\left(\mathbf{R}\right)\hat{\psi}_{e}\left(\mathbf{R}\right)-\hat{\psi}_{g}^{\dagger}\left(\mathbf{R}\right)\hat{\psi}_{g}\left(\mathbf{R}\right)\right]\notag,\\
&&\hat{H}_{ int}=\frac{4\pi \hbar^2a_{eg}^{-}}{m}\int  d^3\mathbf{R}\hat{N}_{e}\left(\mathbf{R}\right)\hat{N}_{g}\left(\mathbf{R}\right)+ \sum_{\alpha\beta}\frac{3\pi\hbar^2b^3_{\alpha\beta}}{m}\int d^3\mathbf{R} \times \notag \\
&&\left[ \hat{\psi}^{\dagger}_{\beta}(\mathbf{R})\left(\nabla \hat{\psi}_{\alpha}(\mathbf{R})\right)-\left(\nabla \hat{\psi}_{\beta}(\mathbf{R})\right)\hat{\psi}_{\alpha}(\mathbf{R})\right]^\dagger \cdot\left[ \hat{\psi}^{\dagger}_{\beta}(\mathbf{R})\left(\nabla \hat{\psi}_{\alpha}(\mathbf{R})\right)-\left(\nabla \hat{\psi}_{\beta}(\mathbf{R})\right)\hat{\psi}_{\alpha}(\mathbf{R})\right]. \notag
\end{eqnarray}}Here the clock  laser (detuned from the bare atomic transition by $\delta=\omega_c-\omega_0$ and with Rabi frequency $\Omega$) propagates  along the axial direction, $\hat Z$, with wavevector  ${ k_c}= 2\pi /\lambda_c$. Atoms are trapped in an external potential $V_{ext}(\mathbf{R})$, generated by the magic lattice beams also propagating along $\hat{Z}$. The lattice induces a weak harmonic radial (transverse) confinement with an angular frequency $ 2 \pi \nu_r$ and creates an array of coupled two-dimensional pancakes. The operator $\hat \psi_{\alpha = g,e }(\mathbf{R})$ is a fermionic field operator at position $\mathbf{R}$ for atoms with mass $m$  in electronic state  $\alpha$, and $\hat{N}_{\alpha}(\mathbf{R})=\hat{\psi}_{\alpha}^{\dagger}(\mathbf{R})\hat{\psi}_{\alpha}(\mathbf{R})$. We have included only $s$-wave and $p$-wave channels, an assumption valid at $\mu$K temperatures. Since nuclear spin-polarized fermions are in a symmetric nuclear-spin state, their $s$-wave interactions are characterized by only one scattering length $a_{eg}^-$,  describing collisions between two atoms in the antisymmetric electronic state, $\frac{1}{\sqrt{2}}(|ge\rangle-|eg\rangle)$.
The $p$-wave interactions enter with three different scattering volumes $b_{gg}^3$, $b_{ee}^3$, and $b_{eg}^3$, associated with the three possible electronic symmetric states ($|gg\rangle$, $|ee\rangle$, and $\frac{1}{\sqrt{2}}(|ge\rangle+|eg\rangle)$, respectively. In addition to elastic interactions, $^{87}$Sr atoms exhibit inelastic collisions. Among those, however, only the $e-e$ ones have been observed to give rise to measurable losses\cite{BishofInelastic}. We denote the  relevant inelastic  $p$-wave scattering length as $\beta_{ee}$.  The magnitude of the measured  $s$- and $p$-wave scattering lengths\cite{XiboSUN} are $a_{eg}^- \sim  68  a_0$, $b_{gg}\sim 74.6 a_0$, $b_{eg}\sim -169 a_0 $
 $b_{ee}\sim -119 a_0$  and  $\beta_{ee}\sim 121 a_0$, with $a_0$ the Bohr radius.


We expand the field operator in  terms of single-particle eigenstates of $\hat{H}_0$, which to a good approximation are harmonic oscillator states along the transverse directions and Bloch functions, $\psi_q(Z)$, along the axial lattice direction. The harmonic oscillator states are characterized by the quantum numbers $ {\mathbf n_r}=(n_X, n_Y)$ and the Bloch functions by the quasimomentum $q$ and band quantum number $n_Z$, which is prepared in only the lowest band $n_Z=0$ for the current loading conditions.

As described in the main text, under our typical operating conditions, the interaction energy per particle is weaker than the spacing between single-particle energy levels. Thus, at the leading order, collisions conserve the total single-particle energy and the atom population is frozen in the initially populated modes which act as effective lattice sites in single particle mode space\cite{ManyBodyTheory}. For an initial state with at most one atom per mode ($|g\rangle$-polarized state),  it is thus possible to reduce  $\hat H$  to a spin-$1/2$ model written in terms of pseudo-spin 1/2 operators   ${\bf \hat S}_{i}= 1/2\sum_{\alpha,\beta}\hat{c}^\dag_{\alpha,q_i, {\bf  n}_{{\bf r}_i}}\vec{\sigma}_{\alpha \beta}\hat{c}_{\beta,q_i, {\bf  n}_{{\bf r}_i} }$. Here $\vec{\sigma}=\left\{\hat \sigma^{x},\hat \sigma^{y},\hat \sigma^{z}\right\}$ are Pauli matrices in the $e,g$ basis and $\hat{c}_{\alpha,q_i, {\bf  n}_{{\bf r}_i}}$ ($\hat{c}^{\dagger}_{\alpha,q_i, {\bf  n}_{{\bf r}_i}}$) are the fermionic annihilation (creation) operators of an atom in the electronic clock state $\alpha$, radial mode ${\bf  n}_{{\bf r}_i}$ and quasimomentum $q_i$.

While the effective spin-spin coupling constants depend on the radial mode quantum  number\cite{MJMManyBodySr, ManyBodyTheory,WallSOC}, to a good approximation, we can replace them by  their  thermally averaged values. Under these approximations we obtain the interaction Hamiltonian $\hat{H}_{\rm int}$ given by Eq.~\ref{eq:IntHamiltonian} in the main text which has been   written in terms  of  the collective spin operators $ \hat S^{x,y,z}$. The $s$- and $p$-wave interactions are given by $U_{eg}^-\approx  \frac{8 \pi \hbar^2}{ma_r^2} \frac{0.09 a_{eg}^-}{aT} \langle W\rangle$, $V_{\alpha \beta} \approx  \frac{12 \pi \hbar^2}{ma_r^2} \frac{0.36 b^3_{\alpha \beta}}{aa_r^2}  \langle W\rangle$
where $\langle W \rangle = a \int dz \vert w_0(Z)\vert^4$ with $w_0(Z)=\frac{1}{\sqrt L} \sum_{q \in BZ} e^{-i q Z} \psi_q(Z)$ the Wannier function for the ground band of the 1D lattice along $Z$ and the summation is over the quasimomenta in the 1st Brillouin zone. Here $a_r=\sqrt{\frac{\hbar}{ 2\pi m \nu_r}}$ the radial harmonic oscillator length, $a_r\approx 450$~nm,  and $ T$  the radial temperature in the effective harmonic oscillator units $h \nu_r /k_B$. For typical experimental conditions, $ T\approx 100$. Extended Data Fig. 2(a) shows the ratio of $\xi/\chi$ as a function of the radial temperature and  Extended Data Fig. 2(b) shows the different interaction parameters as a function of $\nu_{z}$ for different temperatures.  We note that for current experimental conditions, the $s$-wave and $p$-wave interactions are of the same order of magnitude and further cooling of the radial modes would be required to enter the regime where $s$-wave interactions truly dominate.

\begin{figure*}
\centering
\label{fig:Interaction}
  \includegraphics[scale=0.5]{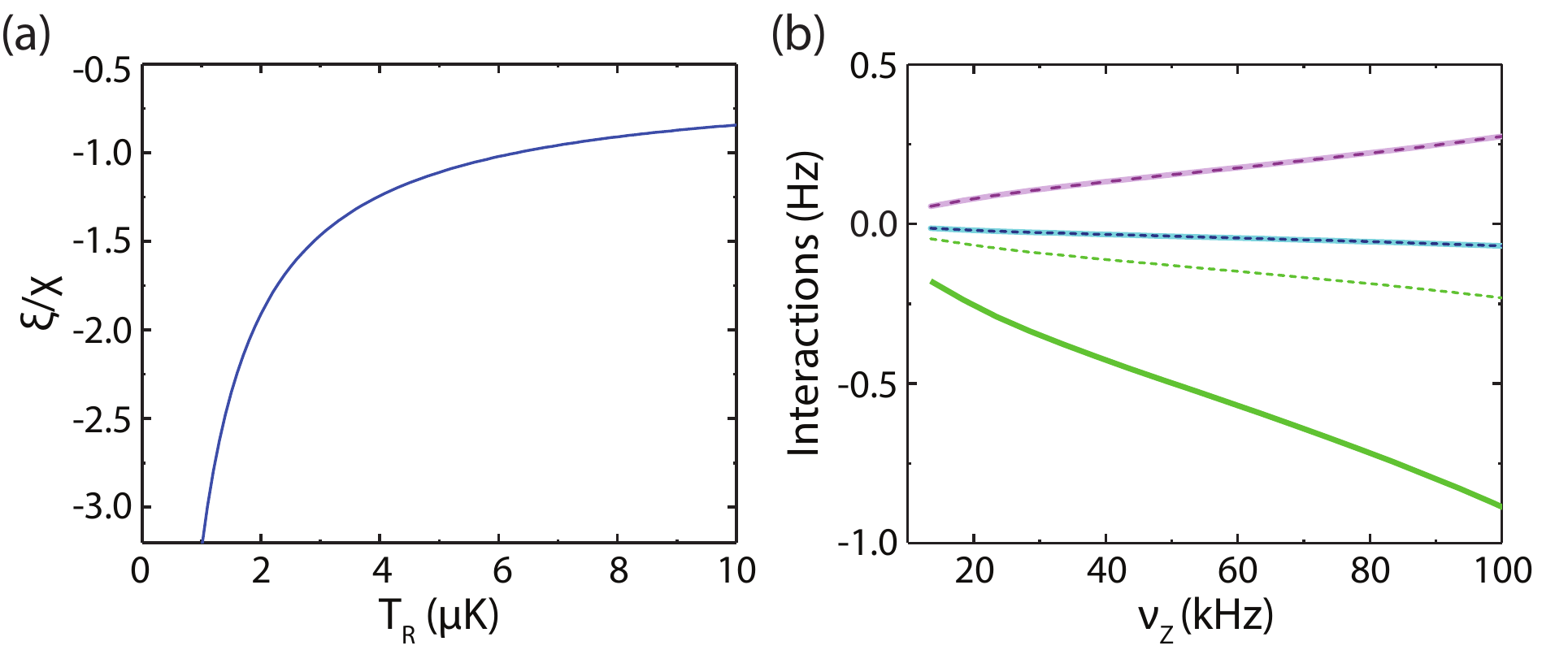}
 \caption*{\textbf{Extended Data Figure 2: Dependence of interactions on temperature and lattice depth} \textbf{(a)} Under current experimental temperatures the exchange interactions ($\xi$) are of the same order as the Ising interactions ($\chi$). \textbf{(b)} Interaction strengths at $T=$ 1$\mu$K (solid lines) and $T=$ 10$\mu$K (dashed lines) for the interaction parameters $\chi$ (purple), $\xi$ (green), and $C$ (Blue) at different lattice depths.  The strong dependence of the exchange interactions on temperature is due to the $s$-wave contribution. The $p$-wave interactions which enter $C$ and $\chi$ are independent of temperature.}
\end{figure*}


\subsection{Mean-field}
For the currently  accessible time scales, beyond mean-field corrections are not resolvable in the dynamics. We have verified this by comparing the mean-field dynamics to the many-body dynamics obtained by the discrete truncated Wigner Approximation (DTWA)\cite{Schachenmayer2015} which accounts for the lowest order quantum correlations. Therefore, it is sufficient to use the mean-field equations of motion to study the dynamics of our system. At the mean-field level, expectation values of products of spin operators are factorized as $\langle \hat{S}^\alpha_i \hat{S}^\beta_j \rangle \approx { \mathcal S}_{i}^\alpha { \mathcal S}_{j}^\beta$, ignoring the build-up of quantum correlations. Here, we have defined  $\langle \hat{S}^\alpha_i\rangle = { \mathcal S}_{i}^\alpha$. The above approximation is consistent with the picture that the net effect of interactions on  an atom $i$ is to induce an effective magnetic field generated by the other atoms. Under the  mean-field approximation, the  Hamiltonian becomes  Eq. \ref{eq:MFHamilton} given in the main text. 

This  Hamiltonian formulation is only valid if one ignores inelastic collisions. To incorporate  them one should formally  use a master equation. However, at the mean-field level the so called recycling terms in the master equation vanish and the inelastic dynamics can be accounted for  by replacing\cite{ManyBodyTheory} $V_{ee}\to V_{ee}-i \Gamma^{ee}/2$ in  Eq. \ref{eq:MFHamilton}, where $\Gamma^{ee} \approx  \frac{12 \pi \hbar^2}{ma_r^2} \frac{0.36\beta^3_{ee}}{aa_r^2}  \langle W\rangle$.  The mean-field equations including the inelastic losses and the single particle SOC terms are given by:
\begin{align}
-i \dot { \mathcal S}_{j}^+ &= \Delta(q_j, \mathbf n_{\mathbf r}) \mathcal S_{j}^+ -2 \frac{\xi}{L} \mathcal S^+ \mathcal S_{j}^z  + \left (2 \frac{\xi +\chi}{L}\mathcal S^z+  \frac{ N C}{L}\right)  \mathcal S_{j}^++ i\frac{\Gamma^{ee}}{2 L}{\mathcal N}^e  \mathcal S_{j}^+ , \\
\dot{ \mathcal S}_{j}^z &=  2 \frac{\xi}{L} \left(\mathcal S_{j}^y \mathcal S^x-\mathcal S_{j}^x \mathcal S^y\right) - \frac{\Gamma^{ee}}{2 L}\mathcal N_{j}^e \mathcal N^e\quad\quad ,\\ 
\dot {\mathcal N}_{j} &= -\frac{\Gamma^{ee}}{ L} \mathcal N_{j}^e \mathcal N^e,
\end{align} where we have used ${ \mathcal S}_{j}^+={\mathcal S}_{j}^x+ {\rm i}{\mathcal S}_{j}^y$, and   $\mathcal {\mathcal N}_{j}={\mathcal N}_{j}^e+{\mathcal N}_{j}^g$ the number of particles in lattice site $j$. Finally $\mathcal N^e=\sum_j \mathcal N_j^e$ and $\mathcal S^\alpha=\sum_j \mathcal S_j^\alpha$ with $\alpha={x, y, z}$. 

We note that within the simulations we include the thermal effects in the single particle dynamics by sampling radial modes for each particle from a thermal distribution. Each particle is assigned  a given $\Delta(q_i, \mathbf n_{\mathbf r})$ depending on the radial mode $\mathbf n_{\mathbf r}$ it occupies. At lowest order in the radial-axial coupling, the radial mode dependence of the  tunneling is given  by\cite{WallSOC,Kolkowitz2017} $J_{\mathbf n_{\mathbf r}}=J_0 +\Delta J (n_X+n_Y),$ with $J_0\approx 1.36 \left(\frac{U_z}{E_R}\right)^{1.057} \exp\left(-2.12 \sqrt{U_z/E_R}\right)$ and
$\Delta J \approx 0.5 \left(\frac{U_z}{E_R}\right)^{1.7} \exp\left(-2.71 \sqrt{U_z/E_R}\right)$.

\begin{figure*}
\centering
\label{fig:SpinLocking}
  \includegraphics[scale=0.5]{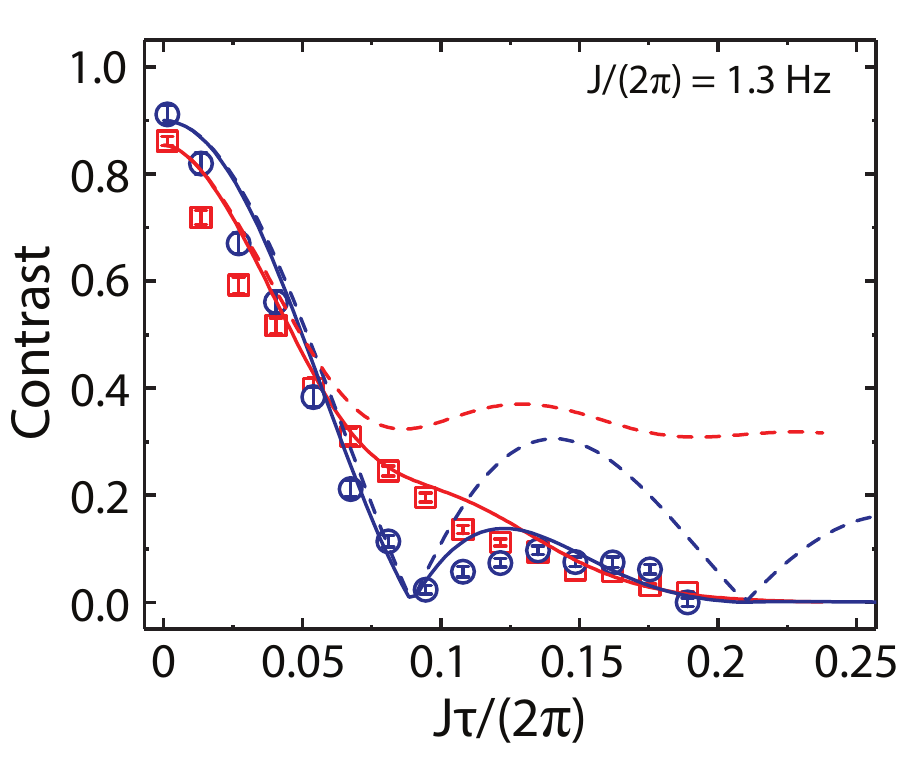}
 \caption*{\textbf{Extended Data Figure 3: Spin Locking without diffusive dephasing.} Same data and theory lines as in Fig.\ref{fig:Contrast}(d) in the main text also showing the predicted contrast in the absence of diffusive dephasing without interactions (blue, dashed) and with interactions (red,dashed). }
\end{figure*}

\subsection{Two Particles}

We can gain insight into the behavior of the many-body system by considering the case of two particles in two sites with  quasimomenta $q_1=0$, and $q_2=\pi$, corresponding to $\Delta(q_i, \mathbf n_{\mathbf r_i})=\pm 4J$, and later extending it to two groups of particles at the aforementioned quasimomenta.  Given that for the parameter regime of interest  the Ising terms  provide a simple collective rotation whose behavior is well understood, as shown in Fig.~\ref{fig:Shift} (a) in the main text,  we are going to ignore them for the following discussion. Namely we will set $\chi=C=0$ and keep only the terms proportional to $\xi$. In this case, we can provide simple analytical expressions for the dynamics:
\begin{eqnarray}
 \label{eq:sx2p} \langle \hat S^x (\tau)\rangle&=&\sin(\theta_1)\left (\frac{2 \xi}{ J_{\rm eff}} \sin \left(\frac{J_{\rm eff}}{2} \tau \right) \sin (\tau \xi  )+ \cos \left(\frac{J_{\rm eff}}{2} \tau\right) \cos (\tau \xi )\right),\\
  \label{eq:sy2p} \langle \hat S^y (\tau)\rangle&=&
\sin(\theta_1)\cos (\theta_1)\left(\frac{2 \xi}{ J_{\rm eff}} \sin \left(\frac{J_{\rm eff}}{2}\tau\right) \cos (\tau \xi ) - \cos \left(\frac{J_{\rm eff}}{2} \tau \right)  \sin (\tau \xi )\right),
 \end{eqnarray}
where $J_{\rm eff}= 2 \sqrt{ (4 J)^2 +\xi^2}$.
From the above expressions, we obtain $\Delta\nu(\tau)=\xi \cos\theta_1\left(\frac{\tan(4 J\tau)}{4 J\tau} -1 \right)$ to first order in interactions. From the above  expressions  the following points are clear: (1) If $\xi=0$ then  $\langle \hat S_y(\tau)\rangle=0$ at any tipping angle and all times and thus  no density shift is present. (2) For   $\xi=0$, at times  $4 J\tau_n=  (2n+1)\pi/2$  the contrast $\mathcal C(\tau_n)$ vanishes  since at those  times  $\langle \hat S_x(\tau_n)\rangle=0$. (3) Finite $\xi$ generates non zero  $|\langle \hat S_y(\tau)\rangle|>0$, introducing a density  shift  for $|\cos(\theta_1) |>0$. The density shifts diverges at linear order of interaction at $\tau_n$. (4)   Finite interactions slow down   the contrast decay. When interactions are weak, $\xi<J$, the contrast decays but the first zero crossing is delayed to later times  $\tau_0\to  \pi( 1+\epsilon)/(8J$) where $\epsilon=(\xi/8J)^2$. Moreover, in this regime, the second revival peak is reduced by  $2 \pi^2 \epsilon$. For large  interactions, $\xi\gg J$, the contrast  no longer decays to zero but saturates at a finite value which approaches  its original value $ \langle \hat S^x (\tau)\rangle\to \sin(\theta_1)$ in the strongly interacting limit. While the conclusions were inferred from the  two particles dynamics they remain approximately valid for the many-body system.   This protection can be seen in Extended Data Fig. 3 where the dashed line displays this saturation for $N/L\sim25$.  We note that the results presented in the main text include the effects of single particle diffusion which hinders our ability to see this phenomena. In particular, for the two  groups of $N/2$ particles, in many situations  one can obtain the approximate $N$ particle dynamics  by replacing  $\xi \to  N\xi/L$. See for example  Eq.~\eqref{eq:SPShift} in the main text.

So far we have mainly discussed the Hamiltonian dynamics.
The immediate effect of the losses is the decay of the $e$ state population,  which modifies  $\mathcal S^z $ and reduces the overall coherence of the state. For $\theta_1=\pi/2$, losses tilt the collective spin out of the equatorial plane and  generate a non-zero $\mathcal  S^y $ component. This in turn helps prevent the contrast from decaying when normalized by the total particle number.

\end{document}